\documentclass[11pt,twoside]{article}

\usepackage{asp2006}
\usepackage{epsf}

\usepackage{natbib}
\def\THCO {$\mathrm{^{13}CO}$} 
\def\NHTD {$\mathrm{NH_2D}$}
\def\AMM {$\mathrm{NH_3}$}
\def\HTCO {\hbox{${\rm H}_2{\rm CO}$}}   
\def\WAT {\hbox{${\rm H}_2{\rm O}$}}   
\def\METH {{${\rm CH}_3{\rm OH}$}}  

\def\HII{H{\sc ii}}

\def\solmass {$\hbox{M}_\odot$}
\def\solum {$\hbox{L}_\odot$}

\def\percc {$\hbox{{\rm cm}}^{-3}$}    

\begin{document}
\title{Initial conditions for massive star formation}   
\author{Friedrich Wyrowski}   
\affil{Max-Planck-Institut f\"ur Radioastronomie, Bonn, Germany}    

\begin{abstract} 
In this contribution, our knowledge of the initial conditions under
which massive star formation takes place is reviewed. Massive stars
are born in massive clumps of giant molecular clouds (GMCs), hence
first the properties of GMCs are summarized. As a potentially early
stage of molecular clouds, infrared dark clouds have been discovered a
decade ago as dark patches in mid-infrared (MIR) images of the
Galactic plane and many studies of the physical conditions within them
have been conducted recently. Without the guidance of MIR absorption,
large scale, unbiased cold dust surveys can be used as well to
identify massive cold clumps. In the absence of indicators of ongoing
massive star formation, like compact HII regions and bright IR
sources, these clumps are the most promising objects for the study of
the initial conditions of massive star formation. Current
observational approaches to find IR quiet clumps and their physical
and chemical properties are summarized.
\end{abstract}



\section{Introduction}

In the last decade, much progress has been made towards understanding
of the earliest phases of massive star formation.  While in the
nineties most studies of early stages targeted hot objects, namely
compact HII regions and hot molecular cores \citep[e.g.][]{kurtz+2000,
churchwell2002}, we learned only in indirect ways, for example through
chemical properties, about their initial conditions.  Time was ripe
then for advances in the field of also colder stages with the advent
of new IR mission, mm/submm bolometer arrays and clever molecular line
observing strategies.  Previously, the initial conditions of massive
star formation were covered in depths in the reviews by
\citet{evans+2002}, \citet{menten+2005}, and \citet{beuther+2007}, but
many new studies were published in the field during the last two
years, justifying a further summary of recent results.

\begin{figure}
\label{fig:l30}
\plotfiddle{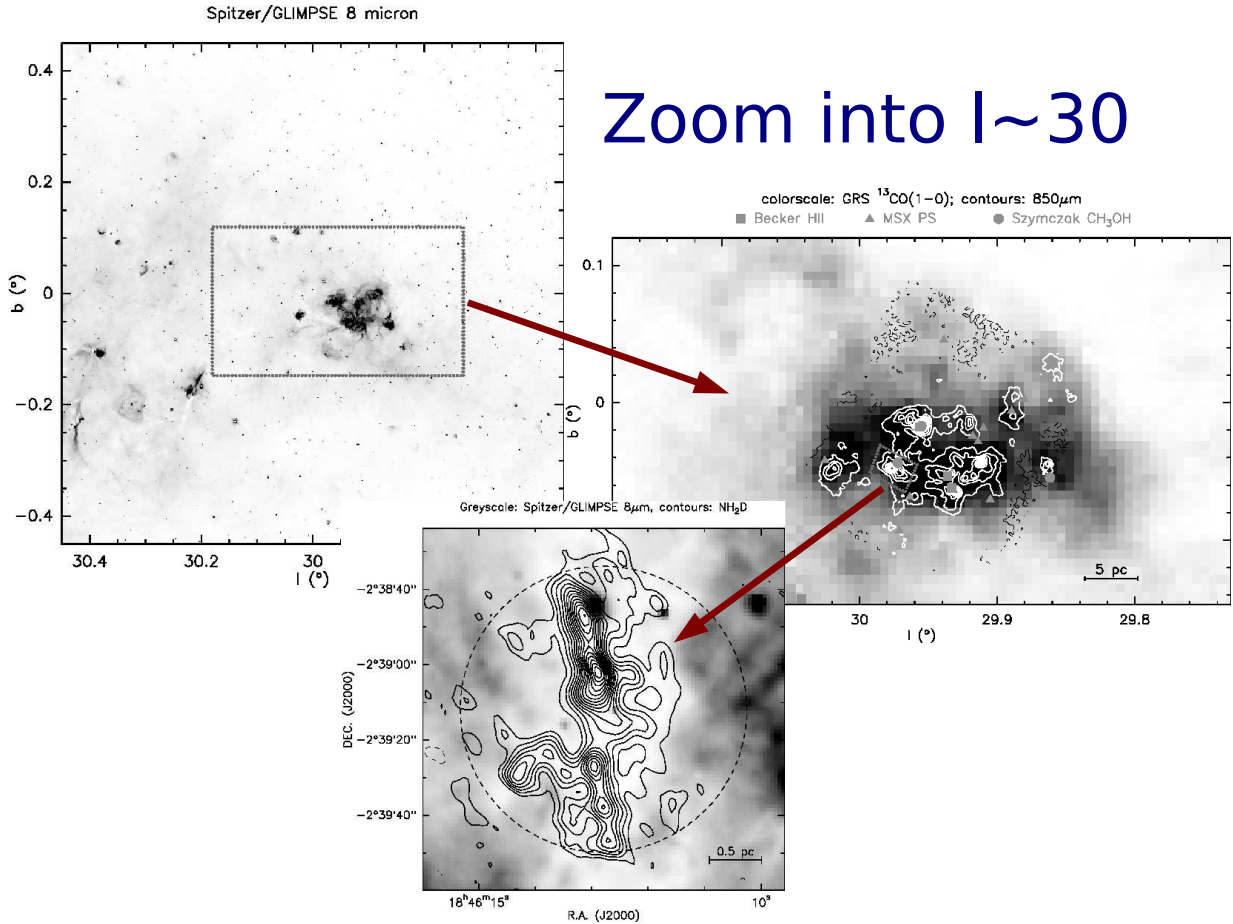}{12cm}{-90}{60}{60}{-240}{360}
\caption{Upper left: Spitzer/GLIMPSE 8~$\mu$m emission of the
         Galactic plane. Upper right: Overlay of submm dust
         continuum \citep{thompson+2005} in contours onto \THCO\ from
         the Galactic ring survey \citep{jackson+2006}. Lower
         left: Overlay of \NHTD\ contours \citep{pillai2006} onto
         GLIMPSE 8~$\mu$m. }
\end{figure}

An instructive example is given in Fig.~\ref{fig:l30}, showing a zoom
into our Milky Way at a Galactic longitude of 30\deg: Massive star
forming regions are among the brightest objects seen in mid-/far-
infrared surveys of the Milky Way. They are distributed along the
Galactic plane with a scale height of only $\sim$~100~pc. Large scale
molecular surveys \citep[e.g. the Galactic Ring
Survey,][]{jackson+2006} show giant molecular clouds (GMCs) with sizes
of tens of parsecs as the birth places of massive star forming
regions.  Submillimeter dust observations of such clouds
\citep[e.g. from the SCAMPS survey,][]{thompson+2005} reveal
parsec-scale dense clumps embedded within the GMCs as the local
environments for different stages of massive star formation, from
relatively late stages like (ultra)compact \HII\ regions (UCHIIs) to
the supposedly earliest stage of massive cold clumps. An
interferometric view of such a region in the cold and dense gas tracer
\NHTD\ is seen in the lower part of Fig.~\ref{fig:l30}, where the
clump breaks up into several highly deuterated 0.1~pc cores, many of
them seen in absorption in the mid-infrared.


This review will mainly cover the observational side of the initial
stages of massive star formation and focus on ``cold'' stages before
infrared-bright massive young stellar objects, hot molecular cores and
UCHII regions form. It is divided into 4 main parts: In
section~\ref{sec:largescale} the large environment will be described,
section~\ref{sec:searches} gives an overview of current searches for
the earliest stages of massive star formation, section~\ref{sec:clumps}
discusses the current knowledge of the properties of the found clumps,
and section~\ref{sec:cores} deals with the small scale view of those
clumps.

\section{Large scale environs}
\label{sec:largescale}

  \subsection{Giant molecular clouds}

From large scale CO surveys
\cite[e.g.][]{heyer+1998,dame+2001,jackson+2006} we know that massive
stars form in giant molecular clouds. These clouds have
diameters of 10--100~pc, masses of $\sim 10^{5-6.5}$~\solmass, mean
densities of several 100~\percc\ but they are strongly clumped
\citep[e.g.][]{stutzki+guesten1990}.


For the substructures in the clouds, different terminologies exists.
In what follows the nomenclature suggested by \citet{williams+2000}
will be used. The condensations seen in the dust continuum map in
Fig.~\ref{fig:l30} are called clumps. In molecular line maps, they are
coherent, overdense structures in l-b-v, which might form whole
clusters. At the distances to the bulk of the massive star forming
regions, at several kpc, they are typical single dish mm-telescope
targets.

Substructures within the clumps are cores. They might form individual
stars or multiple systems and for typical distances, interferometers
are needed to resolve them.

Currently there are two competing views on the physical state of GMCs,
which in this contribution can only be discussed briefly.  The first
one sees GMCs as dynamic, transient objects
\citep[e.g.][]{ballesteros+2007}.  The clouds formed by large scale
colliding gas flows and have lifetimes of about the dynamical crossing
time. The second interpretation considers GMCs as quasi-equilibrium
self-gravitating objects \citep{mckee1999} with lifetimes of many
crossing times. See e.g. \citet{mckee+ostriker2007} for more details.

How can we identify the cold and dense parts of GMCs which might lead
us to the initial stages for massive star formation?

  \subsection{Infrared dark clouds}
  \label{sec:irdcs}

In the late nineties, so-called infrared dark clouds (IRDCs) were
recognized as potential sites for cold precursor stages of MSF.  They
were discovered in large scale mid-infrared surveys of the Milky Way
conducted with space telescopes: \citet{perault+1996} reported
``numerous dark features'' in the ISOGAL survey and \citet{egan+1998}
a ``population of dark cores'' observed in the MSX survey. Those
features are due to absorption of the bright, diffuse MIR emission of
the Galactic plane by cold, high column density clouds
($A_V>25$). Detailed reviews of their properties can be found in
\citet{menten+2005} and also in Jackson (this
volume). \citet{menten+2005} discusses the remarkable resemblance in
dust continuum maps of IRDC G11.11 and the Orion molecular
cloud. Hence, there are large IRDCs that resemble the high column
density parts of GMCs, but there is also a large population of smaller
clouds, down to ``IR dark clumps''. Whether clouds show up as IRDCs
depends strongly on the MIR background and the evolutionary state and
geometry of the GMCs. That the clouds indeed show large column
densities of cold molecular material was shown with \HTCO\ and \AMM\
observations by \citet{carey+1998} and \citet{pillai+2006b},
respectively.

\section{Searching for initial stages of MSF}
\label{sec:searches}

In the pre-IRAS era, most of young massive star forming regions were
found by their association with compact \HII\ regions and maser
emission \citep[e.g.][]{churchwell1990}. With the help of the results
from the IRAS satellite, \citet{wood+churchwell1989} were able to
develop IRAS color selection criteria for ultracompact \HII\ regions
(UCHIIs) which paved the way for systematic studies of UCHIIs, their
environments and also their precursors (see next subsection). From the
galaxy-wide statistics of UCHII regions and massive stars,
\citet{wood+churchwell1989} derived a lifetime of $10^5$~years for the
UCHII stage, much longer than what is expected from the expansion of a
Str\"omgren sphere. Possible solutions to this ``lifetime problem''
are discussed in e.g. \citet{churchwell2002}. In molecular line follow
up studies, many hot molecular cores were found associated with the
UCHIIs, which mark a stage before the occurrence of an UCHII, that
might still be quenched by ongoing accretion
\citep{walmsley1995}. Most of the hot cores are also associated with maser
emission.  Another complementary IRAS color approach was followed by
\citet{henning+1990} to select candidates for so-called BN-type
objects.

\subsection{Maser selected surveys}

Water and methanol masers might trace young stellar objects (YSOs)
$<10^5$~years, \METH\ maser in particular probing only massive YSOs,
hence they can be used as search beacons for early phases of massive
star formation. This was done systematically by
\citet{walsh+1997,walsh+1998,walsh+2003} with surveys for \METH\
masers of objects with IRAS colors resembling UCHII regions and
subsequent follow up programs. An unbiased approach was taken by
\citet{szymczak+2002} with their survey for methanol masers over 20
square degrees of the northern Galactic plane. A much deeper blind
survey was carried out by \citet{pandian+2007}.  The most complete
approach in this respect is the currently ongoing methanol maser
survey of the whole Galactic plane \citep{cohen+2007}.
\citet{plume+1997} studied in detail the physical conditions of
massive clumps associated with \WAT\ masers. But for all of these
studies one has to be aware that maser amplification is strongly
beamed, hence geometry dependent and therefore might not lead to a
complete census of massive YSOs.

\subsection{Infrared selected surveys}

An important step forward was the realization, that the
\citet{wood+churchwell1989} IRAS color criterion to select candidate
UCHII regions will also lead to earlier stages if the
sources have no association with free-free continuum observed
in galaxy-wide cm surveys \citep{palla+1991,sridharan+2002}.

Although in subsequent deeper searches for free-free emission from
\HII\ regions a fraction of the selected sources was detected, a large
amount of likely pre-UCHII regions objects could be identified and
studied in more detail \citep{molinari+1996,beuther+2002a,
beuther+2002b}. A similar approach, making use of the point source
catalog from the MSX survey, was taken by \citet{lumsden+2002} and
\citet{hoare+2004}: they developed color criteria for the MSX
observing bands to select a population of red MSX sources. An
advantage over the IRAS selection is the higher angular resolution
(18$^2$ vs.\ 45x240\arcsec$^2$) and sensitivity of the MSX data.

But all of these searches probe already ongoing star formation and
cannot (directly) identify an earlier cold precluster phase.

\subsection{Cold dust surveys}

Further progress was brought by the availability of sensitive
bolometer arrays to probe the cold dust content of potential
massive star forming regions.

As we have seen in the example of Fig.~\ref{fig:l30}, usually
many clumps can be found embedded in GMCs, which might contain
massive star formation at very different stages. Therefore, one
can search in the larger environs of objects found in the surveys
described in the sections above for earlier, colder clumps. In
the following, several of these surveys will be described in
more detail.

A very prolific instrument was the SIMBA 1.2mm bolometer array
\citep{nyman+2001} operated at the SEST telescope: \citet{hill+2005}
targeted a sample of 131 massive star forming regions selected from
methanol maser and ultracompact HII region surveys which resulted in
the detection of about 400 clumps. About one third of the clumps do not
have a MIR detection with MSX, hence are promising examples of massive
cold clumps in a stage before maser, bright IR sources and UCHII
regions develop. \citet{beltran+2006} observed with SIMBA the
extension to the 4th Galactic quadrant of the IRAS color selected
Palla/Molinari sample. In the 245 observed fields, 95 clumps without
any MSX association were found.  In both of these SIMBA surveys, the
IR-quiet clumps are found to be less massive than the clumps with MSF
association, but this result assumes the same temperatures for all of
the clumps. Therefore the mass could still be similar if the IR-quiet
clumps are much colder, which would be similar to the case of their
lower mass prestellar siblings.

IRAS color selected regions with additional bright CS detections by
\citet{bronfman+1996} were observed with SIMBA by
\citet{faundez+2004}. An analysis of 4 massive cold clumps discovered
in the survey was presented by \citet{garay+2004}.  Most of the fields
targeted contain already UCHIIs, so that this survey covers many
relatively high luminosity sources. This is similar to the approach of
the SCAMPS survey \citep{thompson+2005}, which observed UCHIIs from
\citet{thompson+2006} with evidence for secondary dust peaks in
100\arcmin$^2$ fields with SCUBA. Also in the fields targeted in the
\citet{beuther+2002a} study of IRAS color selected, radioquiet objects
(HMPOs), many cold, secondary clumps were found \citep{sridharan+2005}.
Using MSX and IRAS flux density upper limits, they deduced luminosity
upper limits of 100~\solum\ for an average distance of 4~kpc.

A dust emission survey towards the outer Galaxy was conducted by
\citet{klein+2005} towards a IRAS color selected sample. 128 massive
dust clumps were detected and associated with NIR/MIR and Radio
surveys to place them in a tentative evolutionary sequence from cold
pre-cluster cores to star clusters that have already dispersed their
parental clouds.

A promising hunting ground for massive cold cores are the infrared
dark clouds discussed in Sect.~\ref{sec:irdcs}. Dust continuum surveys
towards samples of IRDCs were conducted by \citet{carey+2000},
\citet{teyssier+2002}, and \citet{rathborne+2006}, with the latter one
covering the largest number of sources. Those mm/submm continuum
surveys reveal a remarkable correlation between mid-IR absorption and
optically thin dust emission. A fraction of the detected dust clumps
is on a closer look associated with compact MIR sources, which is an
indication of ongoing star formation even in the IRDCs
\citep[e.g.][]{pillai+2006a}.


\subsection{Unbiased surveys for cold dust}

A powerful tool for a complete census of massive clumps are unbiased
surveys of either whole giant molecular cloud complex or even of the
whole Galactic plane. Covering several square degrees,
\citet{motte+2007} and \citet{munoz+2007} measured the mm dust
emission in the molecular complexes Cygnus X and NGC6357/NGC6334,
respectively. The Cygnus X survey is described in detail in the
contribution by Motte (this volume).

Several Galactic plane surveys are currently underway or planed.
Results from the CSO/Bolocam 1.1mm survey are presented by Bally (this
volume). ATLASGAL, the Galactic plane survey with the 850$\mu$m LABOCA
bolometer camera at the APEX telescope, has finished its first
coverage of the inner Galactic plane (Schuller et al., in
prep.). These surveys reveal thousands of sources, many of them only
detected at mm/submm wavelengths.

\subsection{Summary}

Recent and current dust continuum surveys produce new detections of a
large numbers of cold massive clumps. A summary of some of the surveys
discussed in the last sections is given in Table~\ref{tab:surveys}.
Although the numbers of IR-quiet clumps are already impressive, very
massive cold clumps which might turn into rich OB clusters are still
rare. This is where the Galactic plane surveys are badly needed. The
variety of selection criteria for the surveys yield sources in a
variety of environs: e.g. with and without nearby powerful OB clusters
and with locations ranging throughout the Galaxy.  Some follow ups to
determine their properties are already completed and will be discussed
in the next section, but given the large amount of new data, several
follow ups are still ongoing.

\begin{table}[t]
\caption{Overview of recent dust continuum surveys. The last 2 columns
         give the total number of detected clumps and the number of
         clumps without association with either compact or diffuse MIR
         in the MSX survey, Note that for the latter, different
         surveys used slightly different criteria.  \vspace{3mm}}
\begin{tabular}{llcrrr}
\hline
\hline
Survey  & Bolometer & Size & RMS  & N(tot)  & N(noMSX)  \\
        &           &      &  mJy/bm  & & \\
\hline
\citet{beuther+2002a}& MAMBO & 69x $\sim$20\arcmin$^2$ & 10-15 & 154 & 56 \\
\citet{hill+2006}   & SIMBA & 131x $\sim$60\arcmin$^2$ & 100-150  & 404  & 112   \\
\citet{beltran+2006}& SIMBA & 235x $\sim$100\arcmin$^2$ & 25-40 & 615  &  95 \\ 
\citet{faundez+2004}& SIMBA & 146x $\sim$150\arcmin$^2$& 40    & 321  &  n/a     \\
\citet{klein+2005} & various & 44x $\sim$10\arcmin$^2$& 3-400  & 126  & 23  \\ 
\citet{motte+2007}  & MAMBO  & 3 sq.deg. & 20    & 129  & 93 \\
\citet{munoz+2007}  & SIMBA & 2 sq.deg. & 25    & 347  & n/a  \\
\citet{rathborne+2006}& MAMBO  & 38x $\sim$36\arcmin$^2$ & 10    & 188  & 140   \\
\hline
\end{tabular}
\label{tab:surveys}
\end{table}

\section{Massive cold clump properties}
\label{sec:clumps}

The properties of the newly found massive cold clumps can be
constrained with molecular line follow-up studies and continuum
observations/data from other wavelengths. The latter can usually be
extracted from available IR surveys, in particular to constrain the
spectral energy distribution of the objects, whereas the line
information needs dedicated observing campaigns. Some of the recent
results will be summarized in this section.

  \subsection{Physical conditions}

\citet{garay+2004} analysed 4 IR-quiet clumps found in the
\citet{faundez+2004} survey. In addition to the dust continuum, also
several CS transitions were observed. They found radii, masses and
densities similar to clumps associated with IRAS and/or UCHII sources,
but the temperatures limits they give for the IR-quiet clumps
($<$17~K) are much lower. Two of the sources were covered also in
the studies of \citet{beuther+2005} and \citet{rathborne+2005} which
found evidence for ongoing star formation in the clumps. An
interesting comparison between low- and high-mass clumps was done by
\citet{garay2005} who found similar scaling relations $\Delta v$(R)
and $n$(R) for the clumps but absolute values for the linewidths and
densities considerably higher for the massive
clumps. \citet{pillai+2006b} studied the temperature of clumps in
IRDCs using ammonia and compared the results with similar studies
towards HMPOs and UCHIIs. They found linewidths and temperatures
clearly smaller than for the more evolved objects, which is confirmed
by the results from \citet{sridharan+2005}.  Detailed modeling of
clumps in an IRDC towards W51 was done by \citet{ormel+2005}. The
submm continuum structure as well as the HCO$^+$ excitation was
included in the spherical models that also considered clumpiness and
turbulence. The main finding is that for all three clumps evidence for
heating and an increase of turbulence to the inside was found, even in
the absence of indications for star formation from the infrared.

\subsection{Spectral energy distributions}

Important information about the clumps can be extracted from their
spectral energy distributions, most importantly the emerging
luminosity. Instructive examples are given in \citet{rathborne+2005}
and \citet{beuther+steinacker2007}, which demonstrate that the 24 and
70~$\mu$m flux measurements that the Spitzer/MIPS instrument can
provide are crucial for the determination of luminosities and the
detection of a warmer component within the clumps. For large samples
of sources, the results from the Spitzer/MIPSGAL survey of the
Galactic plane have the potential to allow detailed studies of the
SEDs of massive cold clumps and their luminosity function.

A promising method to reveal some of the coldest clumps is to make use
of the ISOPHOT serendipity survey at 170~$\mu$m and to cross-correlate
it with the IRAS point source catalog to select sources that are
considerably brighter at 170 than at 100~$\mu$m. \citet{birkmann+2006}
presented continuum and molecular line follow up observations of a
clump found with this method and report temperatures of 12~K a mass of
280~\solmass, and spectra with signs for infall. Although these
characteristics make the clump a very interesting candidate for a cold
precluster object, the data show also outflow wings, which provide
even in this object evidence for already ongoing star formation.

\subsection{Infall}

Evidence for infall in the massive cold clump phase is still scarce
and only a few observations of typical infall tracers have been
published so far. The evidence for infall in the clump observed by
\citet{birkmann+2006} was already mentioned, another source with such
infall evidence is G25.38 studied by \citet{wu+2005} who derive an
accretion rate of $3.4\times 10^{-3}$~\solmass/yr, hence much higher
than rates towards low mass Class 0 sources. But infall is certainly
continuing to much later stages: \citet{fuller+2005} find for the
\citet{sridharan+2002} sample of HMPOs accretion rates of $0.2 - 1
\times 10^{-3}$~\solmass/yr and \citet{wu+evans2003} observed a clear
excess of ``blue'' line profiles towards a subset of the
\citet{plume+1997} water maser sample. The infall might still continue
into the UCHII region phase as mid-$J$ CO line observations by
\citet{wyrowski+2006b} indicate.

\subsection{Chemical conditions}

``Chemistry'' is used here in a very broad sense, covering
multi-line/-molecule observations of massive cold clumps and also
depletion and deuteration studies. A variety of mm lines towards 43 IR
dark clumps in the neighborhood of HMPOs is presented in
\citet{beuther+sridharan2007}. Surprisingly, 18 sources have been
clearly detected in SiO, in some cases with very broad line wings.
The SiO emission is a strong indicator of shocks related to outflow
activity from embedded young stellar objects. An even higher SiO
detection rate towards clumps newly found in Cygnus X is reported by
\citet{motte+2007}, showing that star formation sets in very early on
in the evolution of massive clumps. Other indications of early star
formation activity in the \citet{beuther+sridharan2007} study are an
increase of turbulence to the denser inner parts of the clumps,
measured by the increased line widths of line probes with higher
critical densities, and methyl cyanide detections in 14\% of the
sample, which can be a signpost of embedded hot cores. The observed
methanol abundance in the sample is close to the values found towards
low mass clumps.

Since it is known from low mass clumps that in cold and dense gas
depletion and deuterium fractionation should occur, first studies of
CO depletion and deuterated molecules towards massive clumps have been
performed. \citet{fontani+2006} find depletion and deuteration of
N$_2$H$^+$ towards IRAS selected HMPOs but the observations could not
distinguish between cold, dense gas remnant from the HMPO formation or
secondary colder clumps within the beam. Recent SMA interferometer
follow ups of one of the sources show the N$_2$D$^+$ in several
condensations offset from the HMPO \citep[see Fontani et al., this
volume, and][]{fontani+2007}.  \citet{pillai+2007} observed a sample
of 32 massive clumps in IRDCs and the environs of UCHII regions.  They
find CO depletion factors of $\sim$5 and detected deuterated ammonia
in 23 sources, including some of the highest degrees of deuteration
for ammonia reported so far with values for [NH$_2$D]/[NH$_3$] up to
0.6. No trend of depletion or deuteration with the temperature
measured using ammonia was found, which could be an effect of the
relatively large distances to the massive clumps compared to their
lower mass siblings, so that the gas morphology in the beams is much
more complex. These high deuteration fractions confirm the early
evolutionary stage of the clumps.

%
%
%
%

\subsection{Summary}

Massive cold clumps have higher masses and densities in comparison to
their lower mass cousins. Infall seem to start at this stage but
observations indicate that it continuous also through later stages.
Molecular line studies provide evidence for depletion and, for some
sources, deuteration as high as in low mass prestellar clumps. A wide
range of star formation activity is still hiding in the clumps and 
need to be followed up with high angular resolution studies, which
are described in the next section.

\section{Cores in IR-quiet clumps}
\label{sec:cores}

With the help of powerful interferometers in the cm--submm wavelengths
range, it is now possible to probe the small scale structure within
the IR-quiet clumps discussed in the last section.

\citet{wang+2007} present a high resolution study (with 0.1~pc
beamssize) of the IRDC G28.34+0.06 with the VLA in ammonia: the
northern part of the IRDC harbors already a very active region of star
formation with a luminosity of 10$^4$~\solum, water maser, broad lines
and ammonia temperatures above 30~K. The quiescent cores hide in the
filamentary southern part of the IRDC with temperatures below 20~K and
smaller linewidths. The cores in the southern part show a clear trend
of temperature decrease to their inner high column density parts.

The PdB interferometer has been used by \citet{rathborne+2007} to
image four clumps in IRDCs, which break up into 12 cores. More details
are given in the contribution by Jackson (this volume). With the high
PdBI sensitivity, in one of the cores the ``line forest'' signature of
a hot molecular core was found, which represents an important link
between the early stages of star formation in IRDCs and the later hot
core/UCHII phases. Another PdBI study of a clump in an IRDC is
presented in \citet{beuther+2005}. The main core in the clump has a
mass of 180~\solmass. 4.5$\mu$m Spitzer emission, which is known to be
strong towards molecular outflows, is emanating from the core,
indicating the presence of a collimated flow from a young stellar
object embedded in the core. This is supported by the high resolution
molecular data that shows the turbulent linewidths contribution
increasing to the center.

Yet another high resolution view on a cold, massive clump is given by
\citet{wyrowski+2006a,wyrowski+2007}. Close to the extremely line rich
hot molecular core G327.3--0.6, a massive clump seen in absorption in
the MIR is located. ATCA and APEX observations of N$_2$H$^+$ constrain
the physical conditions in the clump to temperatures of 20~K and
densities in excess of $5 \times 10^6$~\percc. With the high angular
resolution provided by ATCA, the clump breaks up into many cores with
sizes of 0.1--0.2~pc. Only one of them is associated with a very red
Spitzer/GLIMPSE source. The virial mass of the cores is much smaller
than the mass derived from their densities and sizes, hence the
objects might be candidates for massive protocluster in the
process of making, unless collapse is prevented by other means,
e.g. magnetic fields.

%
%

\section{Conclusions and outlook}

With a growing number of large scale cold dust emission surveys, used
in combination with IR surveys of the Galactic plane, this is a very
prolific time for the identification of precluster clump candidates.
The properties of clumps are fairly well known now from the follow-up
studies towards the newly found sources.  Still, there are many cases,
where star formation activity can be found associated with the clump
or the embedded cores on a closer look.  On one hand, this shows that
these clumps indeed started to turn into (massive) stars, hence have
led us to initial conditions of massive star formation, on the other
hand it means that pure ``starless'' objects are still rare, objects
that are yet unaltered by the influence of the star formation process.

For better statistics, unbiased large scale surveys are on the way in
a variety of wavebands: submillimeter bolometer surveys are imaging
the whole Galactic plane, Spitzer provides a spectacular view onto our
Milky Way in the mid-infrared, and the Herschel Space Observatory in
the not so far future will help to put stronger constraints even on the
far-infrared part, where the SEDs of massive cold clumps peak.
Continuing progress on the interferometer side, culminating with the
advent of the eVLA and ALMA, will allow new high resolutions views
of the properties of the cores out of which massive clusters will form.

\acknowledgements 

I would like to thank the organizers for a stimulating conference on
massive star formation and Thushara Pillai, Frederic Schuller and
Malcolm Walmsley for a critial reading of the manuscript.



\bibliography{references}

\end{document}